\documentstyle[twocolumn,aps]{revtex}
\begin{document}
\draft

\title{Change of shape in the yrast sequence in $^{50}\mbox{Cr}$}

\author{L. Zamick and M. Fayache}
\address{Department of Physics and Astronomy, Rutgers University,
	Piscataway, New Jersey 08855}

\author{D. C. Zheng}
\address{Department of Physics, University of Arizona, Tucson, Arizona 85721}

\date{\today}
\maketitle

\begin{abstract}
In shell model calculations for the yrast even spin states of
$^{50}\mbox{Cr}$, the static quadrupole moments of the low spin
states $2^+_1$ and $4^+_1$ are negative
but those of the high spin states $10^+_1$, $12^+_1$ and $14^+_1$
are positive. Even spin states beyond
the single $j$ shell limit, $J_{\rm max}$=14, again have negative moments.
While the $B({\rm E2})$'s for the $J\rightarrow J\!-\!2$
transitions are strongest along the yrast path for $J\le 8$,
it is found that the transition from the second $J$=10 state
to the first $J$=8 state ($10^+_2 \rightarrow 8^+_1$)
is much stronger than the $10^+_1 \rightarrow 8^+_1$ transition.
We also note that while the $16^+_1 \rightarrow 14^+_1$ transition is weak,
the $16^+_1 \rightarrow 14^+_2$ is quite strong.
\end{abstract}


\section{Introduction}
In this work we wish to show that some interesting behavior concerning nuclear
collective motion can emerge from shell-model calculations.
In particular, we will show that as we go up in even spin along an
yrast path, the static quadrupole moments can undergo several changes of sign.

We will focus on the nucleus $^{50}\mbox{Cr}$.
Many years ago, single $j$ shell calculations were performed on this and
neighboring nuclei by McCullen, Bayman, and Zamick (MBZ) \cite{mbz} and by
Ginocchio and French (GF) \cite{Gino}. In these calculations
all the nucleons were in the $f_{7/2}$ shell and the two-body
matrix elements were taken from experiment. Of particular
interest were the cross-conjugate relationships, i.e., the spectrum
of $^{50}\mbox{Cr}$ in the single $j$ shell limit is identical
to that of $^{46}\mbox{Ti}$ \cite{mbz}. This is due to the fact that
in this limited space, the Hamiltonian is invariant under the interchange
of protons and neutron holes and proton holes and neutrons.
For the nucleus $^{48}\mbox{Ti}$, this transformation takes us back to
$^{48}\mbox{Ti}$ and leads to the signature selection rules \cite{mbz,Zamick}.

More recently, however, E.~Caurier {\it et al.} \cite{Caurier}
have shown that the single $j$ shell is inadequate and that by doing
complete $fp$ shell calculations, especially in $^{48}\mbox{Cr}$,
one gets a behavior which approaches rotational motion.

Experimentally, ``preliminary'' results were presented to us by
J. Cameron \cite{Cameron} at the 12th CAP Summer Institute
of Theoretical Physics (July 1994) for possible very high spin states,
i.e., greater than and equal to the single $j$ shell ($f_{7/2}$)
limit of $J_{\rm max}\!=\!14$.
We are however not at liberty to quote these results.

\section{The Single $j$ Shell Results for $^{50}\mbox{Cr}$}

In the single $j$ shell ($f_{7/2}$) calculation for $^{50}\mbox{Cr}$,
one can get states of all spins up to $J_{\rm max}$=14 \cite{mbz,Gino}.
For $J$=14, there is a unique configuration in which
the four protons couple to angular momentum $J_p$=8
and the six neutrons to $J_n$=6. All yrast even spin states up to $J$=12
have been identified experimentally \cite{data}.
The energies of these even $J$ states from $J$=2 to $J$=12 as given in
Ref.\cite{data} are 0.783, 1.881, 3.167, 4.745, 6.341
and 7.614 MeV, respectively.
The single $j$ MBZ results are 1.100, 2.173, 3.264, 4.982, 6.431 and 8.282
MeV.   The overall agreement is quite good, considering
the simplicity of the model.  It should be noted that the MBZ \cite{mbz}
and GF \cite{Gino} single $j$ results are much better than those of
the ``FPD6'' interaction \cite{fpd6} used here (see Table I).
The reason for this is that the ``FPD6'' interaction is designed to give
good results in an extended $fp$ space calculation, rather than
the $f_{7/2}$ space.
In the older calculations, the act of taking matrix elements from experiment
does succeed to a large extent in mocking up the effects of configuration
mixing. However, we will soon see that even if this empirical procedure
works for energies, it does not work for other quantities such as matrix
elements of one-body operators, the quadrupole operator in particular.

In Table I we show the results of the quadrupole moments $Q$ and
$B({\rm E2})$'s for $J \rightarrow J\!-\!2$ transitions
with the ``FPD6'' interaction \cite{fpd6}.
We list separately the proton and neutron quadrupole moments $Q_p$ and $Q_n$
and the ${\rm E2}$ amplitudes $A_p$ and $A_n$ such that
\begin{equation}
B({\rm E2}: \; J \rightarrow J\!-\!2) = (e_p A_p + e_n A_n)^2 ,
\end{equation}
\begin{equation}
Q =  e_p Q_p + e_n Q_n ,
\end{equation}
where $e_p$ and $e_n$ are the effective charges for protons and neutrons,
which we take to be 1.5 and 0.5.
We note that the $B({\rm E2})$'s along the yrast path
are quite strong but the quadrupole
moments change sign as we go from low spin to high spin.

It is not difficult to understand the single $j$ results for the
quadrupole moments.
For the $f_{7/2}$ shell, the quadrupole moment of a single proton is negative:
\begin{equation}
Q(1p:\, j) = -\frac{2j-1}{2j+2} \langle r^2\rangle = -3b^2 \; .
\end{equation}
The quadrupole moment for a proton hole is opposite in sign to that
of a proton: +$3b^2$. Here
$b = \sqrt{\hbar/(M\omega)}$ is the characteristic length of
the harmonic-oscillator basis. For two protons in the $j$ shell coupled
to a total angular momentum of $J$, the quadrupole moment is
\begin{equation}
Q(2p:\, J) = \frac{2 \langle 2\; J\; 0\; J\; |\; J\; J\rangle}
               {  \langle 2\; j\; 0\; j\; |\; j\; j\rangle}
		U( 2\; j\; J\; j\; |\; j\; J) Q(1p).
\end{equation}
Using $b = 2.0\; {\rm fm}$, we get
\begin{eqnarray}
Q(2p:\, J\!=\!2) &=& 7.84\; e\, {\rm fm}^2 , \nonumber \\
Q(2p:\, J\!=\!4) &=& 1.25\; e\, {\rm fm}^2 , \nonumber \\
Q(2p:\, J\!=\!6) &=&-13.70\; e\, {\rm fm}^2 .
\end{eqnarray}
With four protons we are at midshell and the static
quadrupole moment for any state of the $j^4$ configuration is zero.

For $^{50}\mbox{Cr}$, the simplest state is the one with the
maximum possible $J$ in the single $j$ shell, namely, $J = 14$. The
quadrupole moment for this state is
\begin{eqnarray}
Q(^{50}\mbox{Cr}:\, J\!=\!14) &=& e_p Q(4p:\, J\!=\!8)
				+ e_n Q(2\bar{n}:\, J\!=\!6)
					\nonumber \\
&=& 0 + e_n 13.70 = 6.85\; (e\, {\rm fm}^2).
\end{eqnarray}
This explains the $J\!=\!14$ result in Table I. In the above equation,
we use $2\bar{n}$ to represent two neutron holes.
The fact that for two identical particles (holes),
the lower spins $J$=2 and 4 have a positive (negative)
$Q$ and the higher spin $J$=6 has a negative (positive) $Q$, plus
the fact that four identical nucleons have a zero $Q$ help explain why
one gets corresponding results in $^{50}\mbox{Cr}$.

Let us consider one more example, the $J^{\pi}$=$10^+_1$ state. The MBZ
\cite{mbz} wave function is written as
\begin{equation}
|10^+_1\rangle_{\rm MBZ} = \sum_{J_p,J_n} D^J(J_p,J_n) [J_p, J_n]^J \; ,
\end{equation}
where $D^J(J_p,J_n)$ is the probability amplitude that the protons
couple to $J_p$ and the neutrons couple to $J_n$. The wave function
in detail is
\begin{eqnarray}
|10^+_1\rangle_{\rm MBZ}
&=& 0.42 [6,4]^{10} - 0.46 [4,6]^{10} + 0.71 [4^*,6]^{10} \nonumber \\
& &  + [{\rm small\;\; components}]  \; .
\end{eqnarray}
In the above, the states $|4\rangle$ and $|4^*\rangle$ have seniorities
2 and 4 respectively. The fact that the six neutrons (or two neutron holes)
have a high probability of coupling to angular momentum $J_n$=6 accounts
for the fact that for the $10^+_1$ state, the value of $Q_n$ in the
single $j$ shell limit is large and positive.
Again, in this limit, the four protons will have a zero quadrupole moment.
However, as can be seen from the structure of the wave function, we can get
off diagonal non-zero contributions to $Q_p$.

\section{Allowing Nucleons to Be Excited from the $f_{7/2}$ Shell}

In Tables II and III, we give results for $Q$ and $B({\rm E2})$ obtained
in extended space calculations in which a maximum number of
$t$ nucleons ($t$=1 for Table II and $t$=2 for Table III)
are allowed to be excited from the $f_{7/2}$ shell.
In Table IV, results are shown for selected states obtained for
$t > 2$. These results should be compared
with those in Table I from the single $f_{7/2}$ shell model ($t$=0).

The first thing to notice is that the largest spin $J_{\rm max}$ is
18 for $t$=1 and 20 for $t$=2
while it is 14 in the single $j$ shell ($t$=0).
In a complete $fp$ space calculation ($t$=10), the largest even spin
is 22 and the largest overall spin is 23.

Concerning the energy levels, we note that in the larger space
calculation, they are more spread-out. For example, the
$10^+_1$, $12^+_1$ and $14^+_1$ excitation energies in Table I are 3.50, 3.99
and 5.17 MeV respectively, but when up to two nucleons ($t$=2) are allowed
to be excited (Table III), these energies become 5.45, 6.48, and 8.64 MeV
respectively. There is a tendency for the spectrum to look more collective.

Note that the sign changes for the quadrupole moments of
the $6^+_1$ and $8^+_1$ states as we go from $t$=0 to $t=1$, 2 and 3.
The quadrupole moment $Q$ for the $6^+_1$ state is positive
($7.62\, e\, {\rm fm}^2$) in the single $j$ calculation; it changes sign
and becomes -$4.21\, e\, {\rm fm}^2$ in the $t$=1 calculation;
but in the $t$=2 calculation it again becomes positive
($2.65\, e\, {\rm fm}^2$); it undergoes another change of sign and
becomes -$8.11\, e\, {\rm fm}^2$ in the $t$=3 calculation.
In all these calculations, $Q$ has remained
small in magnitude for the $6^+_1$ state.
For the $8^+_1$ state, $Q$ starts with a small and positive number
($1.97\, e\, {\rm fm}^2$) in the single $j$ calculation and becomes
negative for $t$=1 (-$9.73 \, e\, {\rm fm}^2$)
and becomes even more negative for $t$=2 (-$13.37 \, e\, {\rm fm}^2$)
and for $t$=3 ($-\!20.72 \, e\, {\rm fm}^2$).
The quadrupole moment for $16^+_1$ also starts out positive and
becomes large and negative as the configuration space is enlarged.
The $t$=1, 2 and 4 results for $16^+_1$ are
4.64, -0.26 and -9.88 $e\, {\rm fm}^2$, respectively.

The quadrupole moments have remained negative for the lower spin states
$2^+_1$ and $4^+_1$ and positive for the higher spin states
$10^+_1$, $12^+_1$, and $14^+_1$ in different model spaces.
For even higher spin states, $18^+_1$, $20^+_1$ and $22^+_1$
that only exist in the $t\ge 1$, $t\ge 2$, $t\ge 3$ spaces respectively,
the quadrupole moments are
again negative. The quadrupole moments increase in magnitude
with $t$ in general, again indicating more collectivity in larger spaces.

It is interesting to note that the $B({\rm E2})$'s for the
$J \rightarrow J\!-\!2$ transitions get enhanced for
$J^{\pi}$=$2^+_1$, $4^+_1$, $6^+_1$ and $8^+_1$ but the transition
from $10^+_1$ to $8^+_1$ gets reduced as we go to the larger space.
This strange behavior motivated us to look at other transitions, in
particular, the transition from the $10^+_2$ state to the $8^+_1$
state. While for $t$=0, 1, 2 and 3, the values of $B({\rm E2})$
for $10^+_1\rightarrow 8^+_1$ are 80.1, 40.7, 37.8, and 28.7
$e^2 \, {\rm fm}^4$ respectively, the corresponding values
for $10^+_2\rightarrow 8^+_1$ are 25.1, 86.3, 95.3, and 140.0
$e^2 \, {\rm fm}^4$ respectively. This suggests that we associate the
$10^+_2$ state as a member of the ``ground-state band'' which includes
$0^+_1$, $2^+_1$, $4^+_1$, $6^+_1$ and $8^+_1$. This is
further supported by the fact that for $t$=3, the quadrupole moment
of the $10^+_1$ state is large and positive ($45.7\, e\, {\rm fm}^2$)
while that of the $10^+_2$ state is small and negative
($-4.0\, e\, {\rm fm}^2$). The $8^+_1$ moment is also negative.
We also note that although the quadrupole moment $Q$ for the $6^+_1$
state is small and sometimes positive (for $t=0$ and 2),
the $B({\rm E2})$ for its transition to the $4^+_1$ state
(whose quadrupole moment is large and negative)
is strong and gets enhanced as we go from $t$=0 to $t$=2.

Of special interest is the transition $16^+_1$ to $14^+_1$.
This is a transition from a state which cannot exist in the
single $f_{7/2}$ model space to a state which has the highest
possible angular momentum allowed in the single $f_{7/2}$ shell.
The value of $B({\rm E2})$ for this transition is persistently small:
4.8 $e^2\; {\rm fm}^4$ for $t$=1,
6.4 $e^2\; {\rm fm}^4$ for $t$=2 and
4.5 $e^2\; {\rm fm}^4$ for $t$=4.
However, there is a very strong transition from the $16^+_1$ state to
the second $14^+$ state, $14^+_2$: $B({\rm E2}) = 103.5\, e^2\, {\rm fm}^4$
for $t$=2 and $B({\rm E2}) = 122.8\, e^2\, {\rm fm}^4$ for $t$=4.
Presumably the $14^+_1$ state is mainly of an $f_{7/2}^{10}$
character while the $14^+_2$ state, just like the $16^+_1$ and higher spin
states, require excitations of outside the $f_{7/2}$ shell for its existence.
The smallness of the $B({\rm E2})$ for the transition
$20^+_1 \rightarrow 18^+_1$ (and for $22^+_1 \rightarrow 20^+_1$)
can be explained similarly by
noting that $J=18$ is the highest possible angular momentum for the $t$=1
model space while one has to allow two nucleons to be
excited from the $f_{7/2}$ shell ($t$=2) to get $J$=20.

\section{The Rotational Model}

In the extreme rotational model, all the quadrupole moments and
$B({\rm E2})$'s in a rotational band are given in terms of a
single intrinsic quadrupole moment \cite{Bohr}:
\begin{equation}
Q(J) = \frac{3K^2 - J(J+1)}{(J+1)(2J+3)} Q_K \; ,
\end{equation}
\begin{equation}
B({\rm E2}, J_1 \rightarrow J_2) = \frac{5}{16\pi} Q_K^2
	\langle J_1\; K\; 2\; 0\; |\; J_2\; K\rangle^2\; .
\end{equation}
The ratio $B({\rm E2}, J \rightarrow J\!-\!2)/B({\rm E2}, 2^+ \rightarrow 0^+)$
for a $K$=0 band would have the following values for
$J$=2, 4, 6, 8 and 10: 1, 1.43, 1.62, 1.65 and 1.69.
There is a steady but slow rise as $J$ increases.
The corresponding ratios from Table I (single $f_{7/2}$ shell) for
$J^{\pi}$=$2^+_1$, $4^+_1$, $6^+_1$, $8^+_1$ and $10^+_1$ are
1, 1.36, 1.09, 1.15 and 0.76. In the larger space calculation
($t$=2, Table III), for
$J^{\pi}$=$2^+_1$, $4^+_1$, $6^+_1$, $8^+_1$ and $10^+_2$,
the ratios are 1, 1.45, 0.95, 0.97 and 0.63.
These calculated $B({\rm E2})$'s do not obey a simple rotational
formula with a constant $Q_0$. However, the
$B({\rm E2})$'s are still substantial so we can still speak
of the $0^+_1$ to $8^+_1$ and $10^+_2$ states as being part of a band.

{}From the rotational formulae above, the ratio
$B({\rm E2}, J \rightarrow J\!-\!2)/Q^2(J)$ has a value of 0.242 for
$J$=2 and 0.215 for $J$=4. The $t$=2
results from Table III, using effective charges of $e_p$=1.5 and $e_n$=0.5,
yields for the same ratio of 0.313 for $J$=2 and 0.269 for $J$=4.
The agreement is not too bad. However, for $J\ge 6$ there are enormous
deviations between the simple rotational model and the
$t$=2 shell model so that it makes no sense to compare the two
sets of results.

\section{Very High Spin States -- More Accurate Calculations}
In Table IV, we present results for high spin states with as large as
possible values of $t$ as we can handle. For $J^{\pi} = 14^+$ and $16^+$, we
have $t$=4; for $18^+$, $t$=6 and for $20^+$ and $22^+$, we have
complete $fp$ shell calculations ($t$=10).

We see that for $14^+_1$, the $t$=4 results do not differ so much
from the $t$=2 results. However, when we consider states with
$J > 14$, which we recall do not occur in an
$(f_{7/2}^4)_p (f_{7/2}^6)_n$ configuration, there is
a large difference between the $t$=2 and the higher $t$ results.
For example, when $t$=2, the values of $Q_p$ and $Q_n$ from
Table III for $J^{\pi}=16^+_1$ are -1.37 and 3.58 $e\, {\rm fm}^2$.
For $t$=4, the corresponding values, shown in Table IV,
are -4.77 and -5.47 $e\, {\rm fm}^2$. For
$20^+_1$, the $t$=2 values are -2.24 and -14.15 $e\, {\rm fm}^2$
but for $t$=10 (full $fp$ space) the
values are -14.34 and -20.81 $e\, {\rm fm}^2$.
For $J\ge 16$, it is clearly essential to do the calculation with high values
of $t$. Fortunately, the number of configurations is less for
very high $J$ than it is for low $J$ so that such high $t$ calculations
can be done more readily.
For example, to form a state with $J$=22, at least two protons and
two neutrons have to be in the $f_{7/2}$ shell, therefore
the $t$=10 calculation for this state is equivalent to the $t$=6 calculation.

We can gain insight as to why the very high spin states have
negative quadruple moments by looking at the largest spin, $J$=23.
To form an $M$=23 state, we write down the following configuration:
\begin{eqnarray}
&{\rm Neutrons:} & f_{7/2,+7/2}, f_{7/2,+5/2}, f_{7/2,+3/2}, f_{5/2,+5/2},
						\nonumber \\
&		 & f_{5/2,+3/2}, p_{3/2,+3/2}; \nonumber \\
&{\rm Protons:}  & f_{7/2,+7/2}, f_{7/2,+5/2}, f_{7/2,+3/2}, f_{5/2,+5/2}.
\end{eqnarray}
Note that $M_n$=13 and $M_p$=10.
The neutron configuration is unique. For the protons, there are two
additional configurations with $M_p$=10 which can be obtained
by replacing $f_{7/2,+3/2}$ in the above by $f_{5/2,+3/2}$ or $p_{3/2,+3/2}$.
We choose $f_{7/2,+3/2}$ which has the lowest
single particle energy for simplicity.

With the above wave function, the proton quadrupole moment $Q_p$ is
just the sum of the four single particle moments; likewise for the
neutrons. The results are: $Q_p=-4.714 b^2$ and $Q_n=-6 b^2$.
With $b=2\, {\rm fm}$, we get
$Q_p = -18.86 \, e\, {\rm fm}^2$ and $Q_n=-24 \, e\, {\rm fm}^2$.
The $t$=10 shell model result for $Q_n$ is exactly the same, as
it should be since the neutron configuration is unique.
For $Q_p$, the $t$=10 shell model result is
$Q_p = -23.00 \, e\, {\rm fm}^2$ due to configuration mixing.

\section{Closing Remarks}
We have shown that the shell model is useful for examining and uncovering
non-trivial aspects of nuclear collectivity.
By calculating static quadrupole moments in extended $fp$ spaces
(Table III for $J\le 4$ and Table IV for $J\ge 6$), we
are able to find a change of shape in the yrast band
from prolate at low $J$ ($J \le 8$)
to oblate at higher $J$ ($10^+_1$, $12^+_1$ and $14^+_1$ where
$J$=14 is the maximum spin allowed by the
$f_{7/2}^n$ configuration) and another change of shape
from oblate back to prolate for even higher $J$ ($J\ge 16$).
By looking at $B({\rm E2})$'s, it appears that
the even spin states $0^+_1$, $2^+_1$, $4^+_1$, $6^+_1$, $8^+_1$,
and $10^+_2$ form a prolate band, while $10^+_1$, $12^+_1$, and $14^+_1$
belong to an oblate band. The prolate band member $J^{\pi} = 6^+_1$
has a smaller (in magnitude) quadrupole moment compared to
other members of the band whose quadrupole moments are all large
and negative.

The higher spins ($J\ge 16$)
which are only present in the extended $fp$ space do not appear to be part
of this oblate band because the transition $16^+_1 \rightarrow 14^+_1$
is weak and because their static quadrupole moments are negative while
the members of the oblate band, $10^+_1$, $12^+_1$ and $14^+_1$
have (calculated) positive quadrupole moments.
We note that while the E2 transition
$18^+_1 \rightarrow 16^+_1$ is strong,
the transitions $16^+_1 \rightarrow 14^+_1$, $20^+_1 \rightarrow 18^+_1$
and $22^+_1 \rightarrow 20^+_1$ (see Table IV)
are weak. This can be attributed to the fact that
the initial and final states in these weak transitions
are dominated by different configurations:
$(f_{7/2})^{10}$ for $J$=14,
$(f_{7/2})^9\; (p_{3/2}\, f_{5/2})^1$ for $J$=16 and 18,
$(f_{7/2})^8\; (p_{3/2}\, f_{5/2})^2$ for $J$=20, and
$(f_{7/2})^7\; (p_{3/2}\, f_{5/2})^3$ for $J$=22.

The most credit for the resurgence of interest in this region of the
periodic table ($^{48}\mbox{Cr}$ and environs) goes to
E. Caurier and collaborators \cite{Caurier}
who not only developed a code which made it possible to handle
large model spaces, but also made the important connection between the
shell model and nuclear collectivity. With all the advances up to now,
it still must be acknowledged that the convergence in terms of number
of particles excited is rather slow. To make a precise connection between
the shell model and nuclear collectivity is still not a
trivial matter. New technologies for speeding up the shell
model by the Caltech group \cite{koonin} look very promising.

The old problem of choosing the proper shell model basis to explain
collective motion has not yet been properly solved for the $fp$ shell ---
we recall the early work of Bhatt and McGrory \cite{Bhatt} who
showed that while SU(3) might work in the lower part
of the $sd$ shell, this is not the case in the $fp$ shell.
The culprit is the large $p_{3/2} - f_{7/2}$ splitting.
Thus one can invent truncation schemes which appear to be clever but
which don't really work. Thus the large shell model approach, however
distasteful it may appear to some people, seems to be the best way to go
at the moment.

\acknowledgements
This work was supported in part by a Department of Energy grant
DE-FG05-86ER-40299 (L.Z. and M.F.)
and by a National Science Foundation grant PHY93-21668 (D.C.Z.).


\begin{onecolumn}

\begin{table}

TABLE I. The single $j$ results (i.e., $t$=0) for the excitation
energies $E_x$ (in units of MeV) and the quadrupole moments
$Q$ (in units of $e\, {\rm fm}^2$) of
the lowest one or two even $J$ states in $^{50}\mbox{Cr}$
with the ``FPD6'' interaction \cite{fpd6}.
The $B({\rm E2})$'s (in units of $e^2\, {\rm fm}^4$) for the
relatively strong E2 transitions from these states are also listed.
The $B({\rm E2})$'s for a few weak transitions are also given for
the purpose of comparison.
The effective charges $e_p=1.5$ and $e_n=0.5$
are used to compute $Q$ and $B({\rm E2})$. For other choices of the effective
charges, $Q$ and $B({\rm E2})$ can be obtained from $Q_p$, $Q_n$ and
$A_p$, $A_n$ shown in the table through: $Q = (e_p Q_p + e_n Q_n)$
and $B({\rm E2}) = (e_p A_p + e_n A_n)^2$.
$J\!=\! 14$ is the maximum possible $J$ value in this model space.

\begin{tabular}{crrrr|crrr}
Initial State & $E_x$ & $Q_p$ & $Q_n$ & $Q$ \hspace{0.05in}
	& Final State & $A_p$ & $A_n$ & $B({\rm E2})$ \\ \hline
$2^+_1$ & 0.913   &  -3.24 & -5.86 & -7.79 & $0^+_1$ & 4.58 & 4.02 &  78.8 \\
$4^+_1$ & 1.655   &  -3.35 & -5.76 & -7.90 & $2^+_1$ & 5.50 & 4.17 & 106.9 \\
$6^+_1$ & 2.107   &   5.34 & -0.78 &  7.62 & $4^+_1$ & 4.98 & 3.57 &  85.6 \\
$8^+_1$ & 2.954   &   1.89 & -1.72 &  1.97 & $6^+_1$ & 5.01 & 4.00 &  90.6 \\
$8^+_2$ & 3.573   &   6.72 &  4.62 & 12.39 & $6^+_1$ & 0.29 &-0.29 &   0.1 \\
$10^+_1$& 3.496   &  10.13 &  8.93 & 19.66 & $8^+_1$ & 4.12 & 3.16 &  60.1 \\
        &         &        &       &       & $8^+_2$ & 3.26 & 1.00 &  29.1 \\
$10^+_2$& 3.871   &   4.58 &  5.32 &  9.54 & $8^+_1$ & 2.55 & 2.39 &  25.1 \\
	&	  &	   &	   &	   & $8^+_2$ & 1.39 & 2.67 &  11.7 \\
$12^+_1$& 3.993   &   0.41 & 13.02 &  7.12 & $10^+_1$& 3.83 & 1.97 &  45.3 \\
	&	  &	   &	   &	   & $10^+_2$& 0.96 & 2.27 &   6.6 \\
$12^+_2$& 5.316   &   0.14 & -1.25 & -0.42 & $10^+_1$& 1.58 & 0.71 &   7.4 \\
	&	  &	   &	   &	   & $10^+_2$& 3.04 & 0.87 &  24.9 \\

$14^+_1$& 5.166   &   0.00 & 13.70 &  6.85 & $12^+_1$& 4.68 & 0.81 &  55.0 \\
\end{tabular}

\end{table}

\vspace{0.3in}

\begin{table}

TABLE II. Same as Table I but for an extended model space in
which one nucleon is allowed to leave the $f_{7/2}$ orbit and
occupy the rest of the $fp$ shell (i.e, $t$=1).
$J\!=\! 18$ is the maximum possible
$J$ value in this model space.

\begin{tabular}{crrrr|crrr}
Initial State & $E_x$ & $Q_p$ & $Q_n$ & $Q$ \hspace{0.05in}
	& Final State & $A_p$ & $A_n$ & $B({\rm E2})$ \\ \hline
$ 2^+_1$ & 0.726  & -9.74 &-11.14 &-20.18 & $ 0^+_1$ & 5.67 & 5.87 &130.8 \\
$ 4^+_1$ & 1.555  &-12.36 &-12.22 &-24.64 & $ 2^+_1$ & 7.05 & 5.92 &183.4 \\
$ 6^+_1$ & 2.349  & -1.15 & -4.95 & -4.20 & $ 4^+_1$ & 6.22 & 5.72 &148.6 \\
$ 8^+_1$ & 3.470  & -3.85 & -7.89 & -9.73 & $ 6^+_1$ & 6.13 &-6.04 & 38.2 \\
$ 8^+_2$ & 4.229  &  7.12 & 10.98 & 16.17 & $ 6^+_1$ & 0.56 & 0.28 & 1.0 \\
$10^+_1$ & 4.277  & 16.58 & 17.02 & 33.39 & $ 8^+_1$ & 3.13 & 3.37 & 40.7 \\
         &        &       &       &       & $ 8^+_2$ & 3.56 & 1.09 & 34.6 \\
$10^+_2$ & 4.807  &  2.54 &  2.36 &  4.99 & $ 8^+_1$ & 4.55 & 4.92 & 86.3 \\
	 &	  &	  &	  &	  & $ 8^+_2$ & 0.46 & 1.99 & 2.8 \\
$12^+_1$ & 5.191  &  3.45 & 17.27 & 13.81 & $10^+_1$ & 4.04 & 2.45 & 53.1 \\
	 &	  &	  &	  &	  & $10^+_2$ & 1.83 & 3.12 & 18.5 \\
$12^+_2$ & 6.825  & -2.10 & -1.42 & -3.86 & $10^+_1$ & 1.22 & 0.76 &  4.9 \\
	 &	  &	  &	  &	  & $10^+_2$ & 4.05 & 2.97 & 57.2 \\
$14^+_1$ & 6.973  &  1.78 & 18.39 & 11.87 & $12^+_1$ & 4.84 & 2.02 & 68.5 \\
$14^+_2$ &10.493  & -3.76 & 10.39 & -0.44 & $12^+_2$ & 0.28 & 0.88 & 0.7 \\
$16^+_1$ &11.851  & -0.20 &  9.88 &  4.64 & $14^+_1$ & 0.66 & 2.42 &  4.8 \\
         &        &       &       &       & $14^+_2$ & 0.77 &-0.07 & 1.3 \\
$18^+_1$ &13.634  &  0.00 & -1.71 & -0.86 & $16^+_1$ & 2.33 & 3.42 & 27.1 \\
\end{tabular}

\end{table}

\vspace{0.3in}

\begin{table}

TABLE III. Same as Table I but for an extended model space in
which a maximum of two two nucleons are allowed to leave the
$f_{7/2}$ orbit and occupy the rest of the $fp$ shell ($t$=2).
$J\!=\! 20$ is the maximum possible $J$ value in this model space.

\begin{tabular}{crrrr|crrr}
Initial State & $E_x$ & $Q_p$ & $Q_n$ & $Q$ \hspace{0.05in}
	& Final State & $A_p$ & $A_n$ & $B({\rm E2})$	\\ \hline
 $ 2^+_1$ &  1.051  & -10.62 &-12.14 &-22.01 & $ 0^+_1$ & 6.11 & 6.32 &151.8\\
 $ 4^+_1$ &  2.061  & -14.32 &-14.18 &-28.56 & $ 2^+_1$ & 7.63 & 6.76 &219.6\\
 $ 6^+_1$ &  2.934  &   2.41 & -1.95 &  2.65 & $ 4^+_1$ & 6.11 & 5.67 &144.1\\
 $ 8^+_1$ &  4.405  &  -5.52 &-10.18 &-13.37 & $ 6^+_1$ & 6.04 & 6.19 &147.6\\
 $ 8^+_2$ &  5.312  &   5.59 & 10.95 & 13.86 & $ 6^+_1$ & 1.33 & 0.52 &  5.1\\
 $10^+_1$ &  5.446  &  18.43 & 18.40 & 36.85 & $ 8^+_1$ & 2.98 & 3.36 & 37.8\\
          &         &        &       &       & $ 8^+_2$ & 2.90 & 0.82 & 22.7\\
 $10^+_2$ &  6.015  &   1.44 &  1.28 &  2.80 & $ 8^+_1$ & 4.68 & 5.50 & 95.3\\
          &         &        &       &       & $ 8^+_2$ &-0.03 & 1.67 &  0.6\\
 $12^+_1$ &  6.482  &   4.22 & 17.74 & 15.20 & $10^+_1$ & 3.70 & 2.42 & 45.8\\
          &         &        &       &       & $10^+_2$ & 1.66 & 3.05 & 16.2\\
 $12^+_2$ &  8.389  &  -3.09 & -2.67 & -5.97 & $10^+_1$ & 1.37 & 1.04 &  6.7\\
          &         &        &       &       & $10^+_2$ & 4.26 & 3.43 & 65.8\\
 $14^+_1$ &  8.642  &   1.34 & 17.93 & 10.97 & $12^+_1$ & 4.69 & 1.95 & 64.1\\
 $14^+_2$ & 11.937  &  -0.53 &  4.01 &  1.21 & $12^+_2$ & 0.62 & 1.77 &  3.3\\
 $16^+_1$ & 13.883  &  -1.37 &  3.58 & -0.26 & $14^+_1$ & 0.90 & 2.34 &  6.4\\
          &         &        &       &       & $14^+_2$ & 4.82 & 5.90 &103.5\\
 $18^+_1$ & 14.468  &  -2.45 & -2.39 & -4.87 & $16^+_1$ & 3.85 & 5.03 & 68.6\\
 $20^+_1$ & 21.424  &  -2.24 &-14.65 &-10.69 & $18^+_1$ & 0.16 & 1.01 &  0.6\\
\end{tabular}

\end{table}

\vspace{0.3in}

\begin{table}

TABLE IV. Same as Table I but for larger model spaces
where a maximum number of $t$ (given in the table) nucleons are allowed
to be excited from the $f_{7/2}$ shell. Only states with $J \ge 6$ are
calculated. $J\!=\! 23$ is the maximum possible $J$ value in the full
$fp$ ($t$=10) space.

\begin{tabular}{ccrrr|crrr}
Initial State &  $t$ & $Q_p$  & $Q_n$ & $Q$ \hspace{0.05in}
	& Final State & $A_p$ & $A_n$ & $B({\rm E2})$  \\ \hline
$ 6^+_1$ &   3  &  -3.37 & -6.10 &  -8.11 \\
$ 6^+_2$ &   3  &  19.04 & 14.80 &  35.95 \\
$ 8^+_1$ &   3  &  -9.37 &-13.33 & -20.72 & $ 6^+_1$ & 6.96 & 7.42 &200.0 \\
$ 8^+_2$ &   3  &   5.41 & 11.54 &  13.89 & $ 6^+_1$ & 1.60 & 1.11 &  8.7 \\
$10^+_1$ &   3  &  22.74 & 23.18 &  45.70 & $ 8^+_1$ & 2.55 & 3.06 & 28.7 \\
         &      &        &       &        & $ 8^+_2$ & 2.81 & 1.34 & 23.9 \\
$10^+_2$ &   3  &  -1.92 & -2.19 &  -3.98 & $ 8^+_1$ & 5.62 & 6.80 &140.0 \\
         &      &        &       &        & $ 8^+_2$ &-0.41 & 1.06 &  0.0 \\
${12^+_1}^{a)}$
         &   3  &   5.83 & 19.64 &  18.56 & $10^+_1$ & 3.69 & 2.46 & 45.9 \\
         &      &        &       &        & $10^+_2$ & 2.03 & 3.26 & 21.9 \\
${12^+_2}^{b)}$
         &   3  &  -5.47 & -3.78 & -10.09 & $10^+_1$ & 1.18 & 1.11 & 5.43 \\
         &      &        &       &        & $10^+_2$ & 4.85 & 4.73 & 93.0 \\
$14^+_1$ &   4  &   1.21 & 18.35 &  10.99 & $12^+_1$ & 4.58 & 2.35 & 64.6 \\
$14^+_2$ &   4  &  -5.44 &-5.19  & -10.76 & $12^+_2$ & 2.74 & 3.85 & 36.4 \\
${16^+_1}^{c)}$
         &   4  &  -4.76 & -5.47 &  -9.88 & $14^+_1$ & 0.89 & 1.57 &  4.5 \\
         &      &        &       &        & $14^+_2$ & 5.11 & 6.83 &122.8\\
${18^+_1}^{d)}$
         &   6  &  -4.90 & -5.22 &  -9.96 & $16^+_1$ & 4.92 & 4.97 & 97.2 \\
$20^+_1$ &  10  & -14.34 &-20.81 & -31.92 & $18^+_1$ &0.60 & 1.03 &  2.0 \\
$22^+_1$ &  10  & -24.27 &-27.28 & -50.04 & $20^+_1$ &1.45 & 1.19 &  7.7 \\
\end{tabular}

\noindent
$^{a)}$ The $t$=4 results for the $Q$'s of the $12^+_1$ state are:
	$Q_p =  6.27\,e\,{\rm fm}^2$,
	$Q_n = 19.88\,e\,{\rm fm}^2$ and
	$Q   = 19.35\,e\,{\rm fm}^2$;

\noindent
$^{b)}$ The $t$=4 results for the $Q$'s of the $12^+_2$ state are:
	$Q_p = -8.25\,e\,{\rm fm}^2$,
	$Q_n = -5.58\,e\,{\rm fm}^2$ and
	$Q   =-15.17\,e\,{\rm fm}^2$.

\noindent
$^{c)}$ The $t$=6 results for the $Q$'s of the $16^+_1$ state are:
	$Q_p = -5.45\,e\,{\rm fm}^2$,
	$Q_n = -6.72\,e\,{\rm fm}^2$ and
	$Q   =-11.54\,e\,{\rm fm}^2$.

\noindent
$^{d)}$ The $t$=10 results for the $Q$'s of the $18^+_1$ state are:
	$Q_p = -4.90\,e\,{\rm fm}^2$,
	$Q_n = -5.23\,e\,{\rm fm}^2$ and
	$Q   = -9.97\,e\,{\rm fm}^2$.

\end{table}
\end{onecolumn}

\end{document}